\documentclass[pre,superscriptaddress,showkeys,showpacs,10pt,tightenlines,floatfix,twocolumn]{revtex4-1}%
\usepackage[utf8x]{inputenc}
\usepackage[T1]{fontenc}
\usepackage{amsmath}
\usepackage{amsfonts}
\usepackage{amssymb}
\usepackage[caption=false]{subfig}
\usepackage{graphicx}
\usepackage{gensymb}
\usepackage{xcolor}
\graphicspath{{Images/}}

\newcommand{\br}{\mathbf{r}}

\newcommand{\ie}{{\it i.e.: }}

\begin{document}
	\title{Quasistatic rheology of soft cellular systems using Cellular Potts Model}
	
	\author{François Villemot, Marc Durand}
	\affiliation{Universit\'{e} de Paris, CNRS, UMR 7057, Mati\`{e}re et Syst\`{e}mes Complexes (MSC), F-75006 Paris, France.}
	\email{marc.durand@univ-paris-diderot.fr} 
	\date{\today}

	\begin{abstract}
	Soft cellular systems, such as foams or biological tissues, exhibit highly complex rheological properties, even in the quasistatic regime, that numerical modeling can help to apprehend. We present a numerical implementation of quasistatic strain within the widely used cellular Potts model. The accuracy of the method is tested by simulating the quasistatic strain 2D dry foams, both ordered and disordered.  The implementation of quasistatic strain in CPM allows the investigation of sophisticated interplays between stress-strain relationship and structural changes that take place in cellular systems.
	\end{abstract}
    \maketitle

\section{Introduction}

Soft cellular systems, that encompass foams, emulsions and biological tissues, are constituted of highly deformable -- yet almost incompressible -- units (bubbles, drops, cells), interacting through attractive adhesive interactions and soft steric repulsions. Interface
energy is key to the cohesion and the rigidity of these
systems, sometimes constituted solely of fluids. Under small strains, they behave elastically. Above a yield value, plastic rearrangements (called T1 events) occur, conferring to these systems a complex rheological behavior \cite{tlili_2015}. The relationship between the macroscopic response and the microscopic details, such as packing fraction and structural disorder, is still the subject of intense research activity \cite{Weaire_2008, Dollet_2014, Geminard_2018}. 
Even the simplest case of quasistatic-regime -- in which the structure is at mechanical equilibrium at every time -- is far from being fully understood, in particular in the shear banding phenomenon.
Numerical tools are proved to be extremely useful to investigate the relationship between microscopic details and macroscopic mechanical response \cite{Kabla_2003, cox_shear_2006, Guyot_2019, cantat2018foams, Osborne_2017}. The cellular Potts model (CPM) is one of the standard numerical modeling of multicellular systems, with various applications ranging from foam coarsening to collective behaviors of biological cells. However, because of its lattice-based modeling technique, it has been rarely used to investigate mechanical properties of cellular systems, except for a few exceptions \cite{Jiang_1999, Raufaste_2007}. 

In this paper, we present a rigorous implementation of quasistatic strain within CPM, offering a versatile tool to investigate interplay between mechanical properties and other processes at work in cellular systems, such as coarsening in foams, or cell division and cell death in biological tissues. 
The outline of the paper is as follows: in Section \ref{Method} we introduce the cellular Potts model and show how it can be conveniently extended to simulate cellular systems under quasistatic strain. The method is compared with other existing approach, and extension to higher strain rates is discussed. In Section \ref{Benchmark} we test the proposed method by analyzing the shear strain of 2D dry foams. For a regular hexagonal foam, the shear modulus we obtained numerically agrees with the corresponding theoretical expression \cite{Princen_1983,khan_rheology_1986}. Yield strain is also analyzed. For disordered foams, we study the effect of disorder on the affinity of the displacement field and the shear modulus magnitude, and compare our results with those reported in literature.


\section{Modeling quasistatic strain with CPM \label{Method}}

\subsection{Cellular Potts Model}

Cellular Potts model (CPM), also called the Glazier–Graner–Hogeweg (GGH) model, is one of the most accepted models of a multicellular system.
It is widely used for simulating
cellular systems in various fields of physics or biology, such as coarsening and mechanics of foams \cite{Glazier1990,Jiang_1999}, tissue morphogenesis \cite{Hirashima_2017}, cell sorting \cite{Graner_1992} and collective cell motion in epithelial tissues \cite{Szabo_2010, Kabla_2012}.
The CPM is a lattice-based model in which each cell in the system is given a different label (cell ID), and each lattice site $k$ has a value $\sigma_k$ taken from the list of cell IDs. A given cell is then represented by the subset of lattice sites that have its cell ID. 
A cell type $\tau(\sigma)$ can also be defined for each cellular domain. The CPM makes no assumption on the shape or the connectivity of the cellular domains, these properties are direct consequences of the energy terms. In particular, walls between adjacent cells are allowed to fluctuate, and T1 events happen spontaneously.
The CPM Hamiltonian $\mathcal{H}$ that characterizes 2D soft cellular systems reads  \cite{Graner_1992}:
\begin{equation}
	\mathcal{H}=\sum_{\substack{neighboring \\ sites  \langle k,l \rangle}}J_{\tau,\tau^\prime}\left(1-\delta_{\sigma_k,\sigma_l}\right)+\frac{B}{2A_0}\sum_{\substack{cells \\ i}}\left(A_i-A_0\right)^2.
 \label{eq:potts-energy}
\end{equation}
The first sum in Eq. (\ref{eq:potts-energy}) is carried over neighbouring sites $\langle k,l \rangle$ and represents the boundary energy: each pair of neighbours having unmatching indices determines a boundary and contributes to the boundary energy.
Here, $\sigma_k$ and $\sigma_l$ are the site values of site $k$ and $l$, respectively.
$\tau$ and $\tau^\prime$ are abbreviations for $\tau(\sigma_k)$ and $\tau(\sigma_l)$. 
$J_{\tau,\tau^\prime}\left(=J_{\tau^\prime,\tau}\right)$ is the energy per unit contact length between cell types $\tau$ and $\tau^\prime$. 
The second sum in Eq. (\ref{eq:potts-energy}) represents the compressive energy of the cells. $B$ is the effective 2D bulk modulus of a cell, $A_i$ is the area of cell $i$, and $A_0$ the \textit{nominal area}.

The state of the system is updated via a Monte Carlo algorithm: a lattice site is first selected randomly. A target label is then randomly selected amongst this site's neighboring labels, and the update is accepted or discarded following Metropolis-like rule which preserve the connectivity of cellular domains \cite{Durand_2016}. 
The acceptance probability in this Monte Carlo scheme requires a temperature, which has to be chosen carefully. Setting the temperature to a small value is analogous to performing an energy minimization, which is the usual choice to study the structure of dry foams. Higher temperatures will induce fluctuations in the boundaries between adjacent bubbles. Further increase of the temperature leads to large topological rearrangements and is useful to model biological systems, the simulation temperature reflecting the cellular activity. 

\subsection{Adding quasistatic strain to CPM}
A common way to introduce strain in numerical simulations is by changing the shape of the simulation box. After proper equilibration, static properties of the materials can be measured directly for any given strain.
%
However, such method cannot be used in lattice-based modeling techniques as CPM. Nevertheless, strain can be applied by adding appropriate terms in the Hamiltonian.
This approach has been used by Jiang and Glazier to simulate foams submitted to a time-dependent shear rate \cite{Jiang_1999}. In this study, shear is introduced by adding an energy contribution of the bubble boundaries, either in the bulk or those in contact with the two edges of the sample only, so that updates that move the wall in the direction of the shear are more likely to be accepted. The energy term added to the Hamiltonian is actually an energy gradient, and is kept constant throughout the simulation. In Monte Carlo, an energy gradient is analogous to a stress, so that the simulations are actually performed by applying a constant stress to the foam.
The same method has been used by Raufaste \textit{et al.} to simulate a foam flow around an obstacle \cite{Raufaste_2007}.
One great advantage of this method is that it does not require to wait for mechanical equilibration before incrementing the wall displacements. However, this approach has also a few drawbacks. First, when the strain energy term is applied on every bubble boundary (bulk strain), it overdetermines the displacement field. The case of a regular hexagonal foam is illustrative in this respect: the method implies an affine deformation of the bubble boundaries, which is not compatible with the Plateau's laws \cite{cantat2018foams}. 
As we will discuss in Section \ref{sec:regular_foam}, exact resolution of the deformation field shows that only the midpoints of the films follow affine displacement \cite{Princen_1983,khan_rheology_1986}.
Similarly in \cite{Raufaste_2007}, the added energy term sets the rheological behavior of the foam, resulting in a plug flow of the foam in the channel.
Second, when the applied stress is larger than the yield value, the foam deformation produces a stress that opposes the one applied by the energy term. As a consequence, the actual shear rate of the simulation is not constant, but is the difference between the applied stress and the stress produced by the foam as an elastic response. This makes evaluating the actual shear strain quite difficult (in \cite{Jiang_1999} it is assumed that it is proportional to the number of MCS).

We develop here an alternate method to simulate quasistatic strain within CPM, while avoiding these drawbacks. It first requires the Potts lattice to be non-periodic in one direction, that we choose to be the $y$ direction. This non-periodicity effectively creates a foam encased between two walls. To simulate a given strain created by the movement of these walls, we impose the displacement of the bubbles that touch the edges $y=\pm L/2$, where $L$ is the size of the box in the $y$ direction. This is done by adding the following term to the Hamiltonian: 
\begin{equation}
  \label{eq:harmonic-com}
  \mathcal H_{strain} = \sum_i \frac{k}{2} \delta_{i,e} \left| \br_i - \br^\star_{i} \right|^2,
\end{equation}
where $\br_i$ is the position of the center of mass (c.m.) of bubble $i$, and $\br^\star_{i}$ its the target position. For instance, to simulate a shear strain $\epsilon$ along the $x$ direction, $\br^\star_{i} =\epsilon ~ y_i \mathbf{e}_x$, where $\mathbf{e}_x$ is the unit vector along the $x$ axis.
The Kronecker delta $\delta_{i,e}$ in Eq. \ref{eq:harmonic-com} is there to restrict the application of the strain  to the bubbles in contact with the top and bottom edges of the box (symbolized with index $e$). The displacement and deformation of the bubbles in the bulk result only from the minimization of the total energy.

The constant $k$ acts as a spring coefficient. Higher values will impose a stronger restraint on the position of the c.m.. This value must be high enough so that the c.m. of each bubble stays in the vicinity of its target position, but not so high that it affects the shape of the bubbles.
It can be chosen empirically: starting from a low value, the constant $k$ can be increased until the average position of each c.m. is close enough to its target value (typically within a distance of $1$ pixel).
Alternatively, an estimation of a proper value can be obtained by considering the stress that these harmonic restraints apply on the bubbles. They create a potential energy gradient around the c.m. of bubble $i$, which is equivalent to a force $k \left( \br_i - \br^\star_{i} \right)$. This force is actually the shear stress that is applied to the system in order to induce the shear strain.
In the elastic regime, this stress is proportional to the strain, and depends only on the shear modulus. With an estimation of the shear modulus, we can find the value of $k$ that will induce an average distance between the c.m. and its target position of any arbitrary value, typically chosen to be of the order of a pixel. With this method, $k$ will depend linearly on the imposed shear strain.

The methodology to simulate quasistatic strain is then the following: starting from an unstrained foam, we slowly increase the applied strain $\epsilon$ over multiple simulations. The strain is kept constant over the course of a given simulation, and the final configuration is used as the starting point of the simulation at a higher strain. The increments of $\epsilon$ must remain small in comparison with $\ell/L$ (where $\ell$ is the typical size of a bubble), especially at yield strain and above, to ensure that the structure relaxes in accordance with a true quasistatic regime, \ie the succession of T1 events that would occur in a real foam is reproduced accurately.
When $\epsilon$ is incremented, bubbles at the boundary translate to their final positions over the course of just a few Monte Carlo steps (MCS). However, proper equilibration over the whole sample takes longer, and depends on both the temperature and the system size. We check that equilibrium is reached by monitoring the total energy, and run each simulation until the energy fluctuates around a steady value.

Ensemble averages and relevant quantities can be obtained for each value of the strain. In particular, the energy is used to extract the relevant information about elastic modulus and yield strain. In the elastic regime, the stress is proportional to the strain, so that the strain energy varies quadratically with the strain, with a prefactor that depends only on the size of the system and the effective elastic modulus. For instance, the energy of a 2D medium with surface area $\mathcal A$ under a shear strain $\epsilon$ is
\begin{equation}
  \label{eq:energy-strain}
  \mathcal E(\epsilon) - \mathcal E_0 = \mathcal A \frac{G}{2} \epsilon^2,
\end{equation}
where $\mathcal E_0$ is the energy at zero strain, and $G$ is the 2D shear modulus. 
Yield strain is determined either by tracking the drop of strain energy or counting the frequency of T1 rearrangements.

\subsection{Validity beyond quasistatic regime}
In the quasistatic regime, viscous dissipation plays no role, and typical timescale of T1 rearrangements  \cite{Durand_2006} is much smaller than timescale of strain. Although it is tempting to simulate mechanics of cellular systems beyond quasistatic regime, it must be warned that CPM is a Monte Carlo simulation technique, and as such the kinetics of the relaxation process is determined by the Monte Carlo updating rule. Therefore, the rheological behavior in this regime will depend on the chosen updating rule. CPM traditionally uses Metropolis-like algorithm, because of its ability to mimic overdamped force-velocity behavior \cite{Glazier2007b}. 


\section{Quasistatic shear of 2D foams \label{Benchmark}}

\subsection{Initial state preparation}

We test our method and assess its performances by simulating 2D foams under quasistatic shear strain. 
Although liquid content can be readily incorporated in CPM, we assume the dry foam limit in this study.
We first simulate regular hexagonal foams, for which quasistatic shear deformation, shear modulus and yield strain can be calculated analytically. We then extend our study to polydisperse disordered foams, to check that our method allows us to capture the effect of disorder, and compare our results with those obtained in a previous study \cite{cox_shear_2006} using Surface Evolver, another popular numerical model for cellular systems \cite{Brakke_1992}. 

For the regular hexagons, we use $100$ bubbles on a $10$ by $10$ arrangement. Periodic boundaries are used along the $x$ direction, but not along the $y$ direction, which effectively results in walls at the top and bottom of the simulation box. The initial and target surface area of the bubbles that lie at the boundaries are set to half of the surface area $A_0=1000$ pixels$^2$ of the other bubbles. This is done so that inter-bubble films meet with the boundaries at $90 \degree$ angles, which is expected for a foam at rest. Special care is taken so that the aspect ratio of the simulation box matches the aspect ratio of a regular hexagonal lattice ($2/\sqrt{3}$).

The polydisperse foam is created from a random distribution of points in a square box. A Voronoi tessellation is done on this array of points to generate the starting configuration. Each bubble is given a random target surface area $A_0$, following a normal distribution with an average of $1000$ pixels$^2$ and a standard deviation of $\Delta A =125$ pixels$^2$.
The foam is then equilibrated in consecutive stages: a temperature annealing is first performed with a small compressibility. Lowering the compressibility enhances topological rearrangements, resulting in faster and more thorough equilibration. In a second stage, the compressibility is progressively increased to a more realistic value.
Finally, the foam is sheared along the $x$ axis in both directions up to a strain of $0.1$. This final step ensures that the shear loading that will be performed in the following will not trigger many, if any, T1 events. At the end of this equilibration procedure, the standard deviation of the distribution of side numbers per bubble is found to be $\Delta n=0.48$.

In all our simulations, we set $J=1$ and $B=30$. Note that, as shearing is done at constant volume, the results are not really affected by the compressibility of the cells.

\subsection{Hexagonal foam \label{sec:regular_foam}}


\subsubsection{Displacement field and shear modulus}

\begin{figure}[h]
	\begin{center}
  \includegraphics[width=0.9\columnwidth]{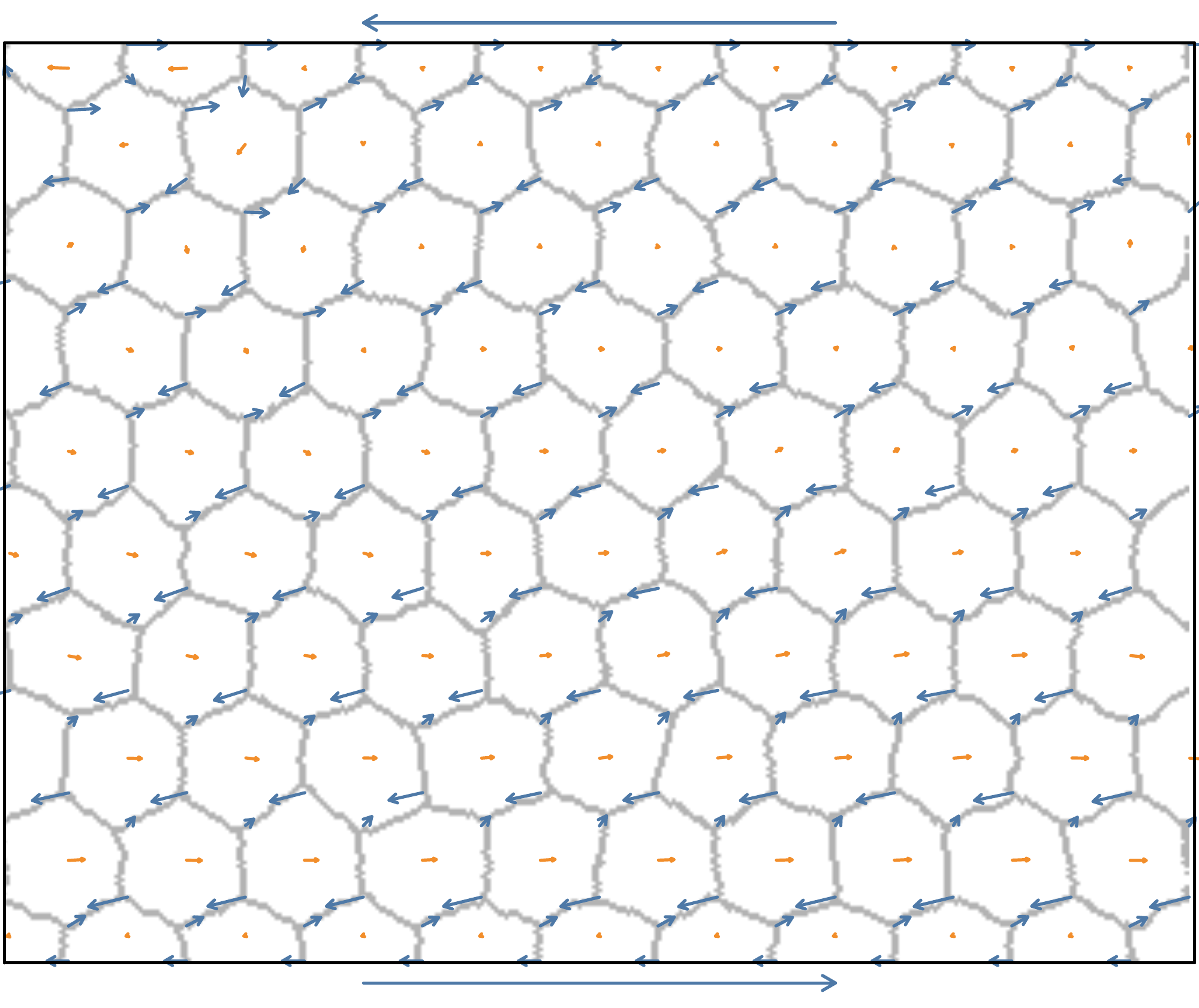}

  \caption{Non-affine component of the displacement field of the bubble centers (in red) and threefold junctions between films (in blue) for a regular hexagonal foam under quasistatic shear (image corresponds to strain value $\epsilon=0.45$).}
  \label{fig:na-hex}
    \end{center}
\end{figure}

For a regular, hexagonal foam, the deformation, and subsequently the elastic moduli can be calculated analytically~\cite{Princen_1983,khan_rheology_1986}: 
since the foam cells are spatially periodic, for any deformation, the centers of the hexagonal cells move affinely with the bulk. Cell symmetry implies that the midpoints of each film also moves affinely. Further, in any deformation, the films remain planar, in accordance with Young and Laplace's equations. Note that, as a consequence, the threefold junctions between films do not move affinely.
We define the non-affine component of the displacement field as the actual displacement field to which the affine displacement field is subtracted. Figure \ref{fig:na-hex} shows the non-affine displacement field of both the bubble centers and the film junctions that we obtained numerically. In agreement with the theory, only bubble centers follow the affine displacement field.

The exact stress-strain relationship has been calculated by Princen, yielding the following expression for the shear modulus~\cite{Princen_1983,khan_rheology_1986}:
\begin{equation}
  \label{eq:stress-strain}
  G_{\text{Princen}} = \frac{2 \lambda}{\sqrt{3} l_h} \frac{1}{\sqrt{\epsilon^2 + 4}},
\end{equation}
where $\lambda$ is the line tension (so a film of length $\ell$ has an energy of $2 \lambda \ell$) and $l_h$ is the length of the side of an hexagon. Equation \ref{eq:stress-strain} gives the shear modulus for a foam up to the yield point. For small strains, its expression simplifies to:
\begin{equation}
  \label{eq:shear-modulus}
  G_{\text{Princen}} = \frac{\lambda}{\sqrt{3}l_h}.
\end{equation}

Note that in CPM simulations, the line tension $\lambda$ is proportional to $J$: $\lambda=zJ$, where the prefactor $z$ depends on the range of interactions between lattice sites \cite{Magno_2015,Villemot_2020}
Actually, the underlying lattice introduces some anisotropy, so that $z$ depends slightly on the orientation of the film. Increasing the neighbor order helps to smooth out this anisotropy, but increases the computational cost. For this reason, we use fourth neighbour order (corresponding to 20 neighbors per pixel) in our simulations, which is a good compromise between cost and accuracy. However, a residual anisotropy can still have significant impact on the mechanical response of the simulated foam, as films tend to be pinned in orientations that minimize energy. This is especially pronounced for the regular hexagonal foam, whose films have three possible orientations only. For disordered foams, anisotropy of the line tension is somehow smoothed out by the wider orientational distribution of the films.

\begin{figure}[h]
  \includegraphics[width=\columnwidth]{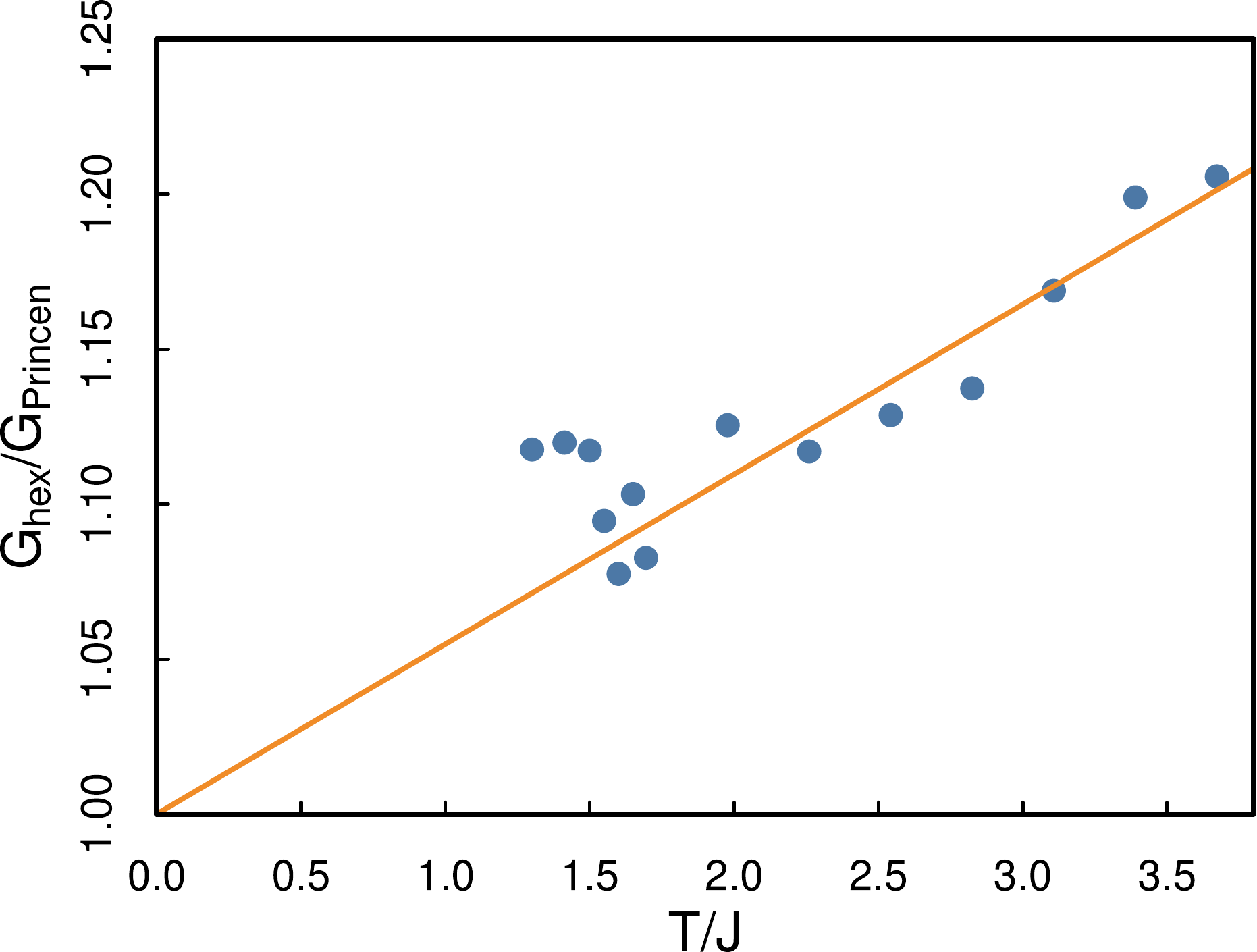}
  \caption{Dimensionless shear modulus $G_{\text{hex}}/G_{\text{Princen}}$ of an hexagonal foam as a function of the reduced temperature $T/J$. The orange line represents the linear fit over the range $T/J > 1.5$.}
  \label{fig:G-T}
\end{figure}

Fortunately, the effect of lattice anisotropy can be circumvented by increasing the simulation temperature, that has for effect to induce sampling over more film orientations. However, increasing the temperature also tends to increase the overall energy of the system, mostly because the fluctuating films are longer than at zero temperature. Precisely, when fluctuations are small, the increase in length of a film at temperature $T$ is proportional to its length at  zero temperature: $\delta \ell = \ell T/2\lambda a$, where $a$ ($a\sim 1$ pixel) is some cut-off length \cite{Villemot_2020}. Therefore, one must evaluate $G$ at different temperature values and then extrapolate its value at zero temperature to circumvent anisotropy artefacts.

For a given simulation temperature, a series of simulations are performed at different shear strains. A quadratic fit of $\mathcal E=f(\epsilon)$  gives us the numerical value of the shear modulus of the hexagonal foam $G_{\text{hex}}$. 
The value of $G_{\text{hex}}$ as a function of the reduced temperature is reported on Fig.~\ref{fig:G-T}.
For temperatures $T/J\leq 1.5$, the energy does not vary quadratically with the strain, because of the anisotropy of the underlying lattice, leading to inaccurate values of $G_{\text{hex}}$. For temperature range  $T/J> 1.5$ on the other hand, the quadratic fit converges and the reported value  $G_{\text{hex}}$ varies linearly with the simulation temperature. We adjust the value of the prefactor $z$ such that the intercept of the linear fit of $G_{\text{hex}}$ is equal to the theoretical value $G_{\text{Princen}}$ (Eq. \ref{eq:shear-modulus}). We obtain  $z=10.50 \pm 0.07$, in very good agreement with other values reported in literature \cite{Kafer_2007, Marmottant_2009, Magno_2015,Villemot_2020}, and hence confirming the accuracy of the method. 

%

\subsubsection{Yield strain}

\begin{figure}[h]
  \centering
  \subfloat[]{
    \includegraphics[width=0.45\columnwidth]{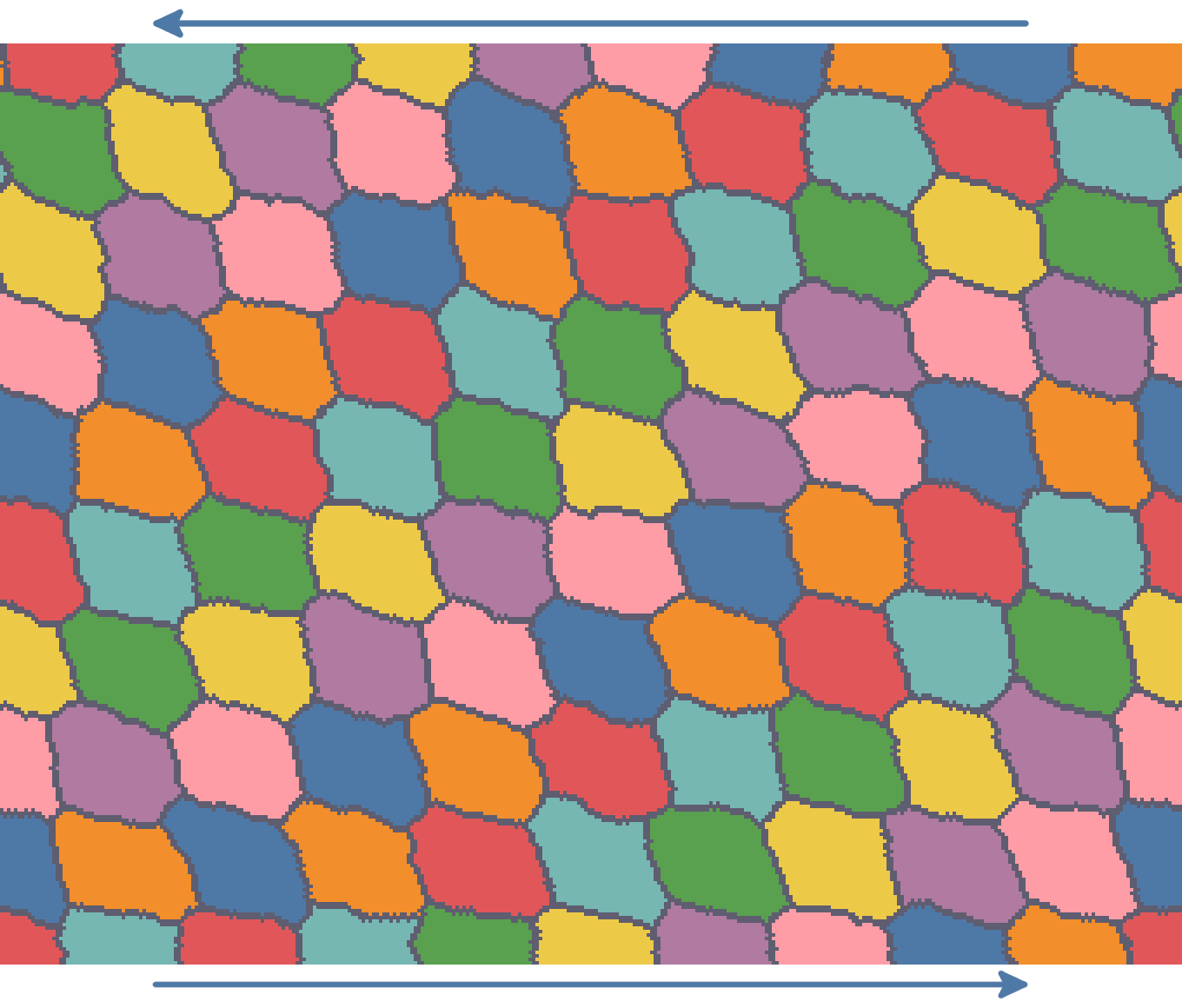}
  } \hfill
  \subfloat[]{
    \includegraphics[width=0.45\columnwidth]{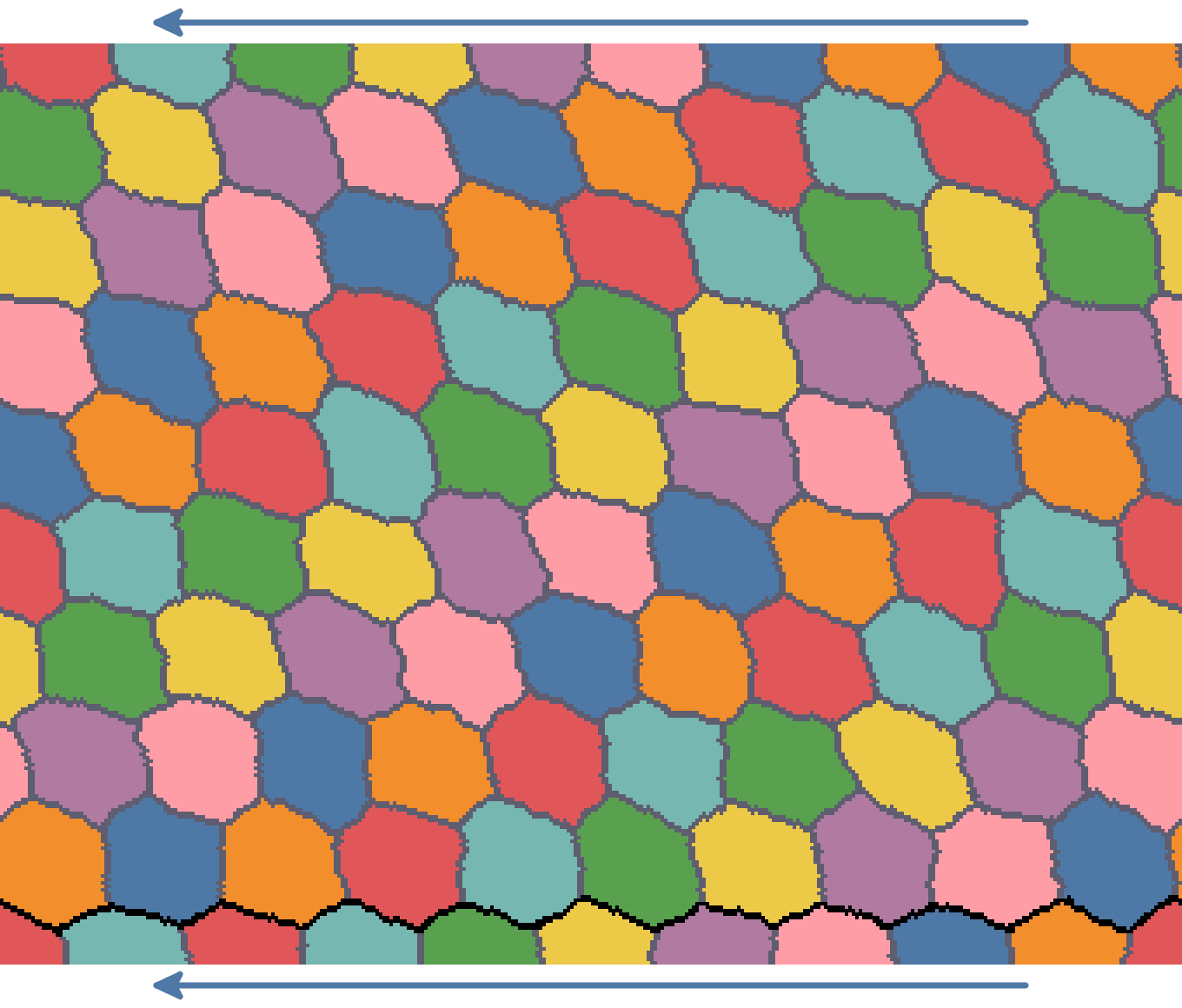}
  } 
  \caption{Configurations right before and after the creation of a shear band. The black line indicates the localization of the shear band.}
  \label{fig:shear-bands}
\end{figure}

Unlike shear modulus, yield strain of a regular hexagonal foam changes with the orientation of strain. With our chosen orientation (Fig. \ref{fig:na-hex}), the expected value is $2/\sqrt{3}$ \cite{Princen_1983, khan_rheology_1986}. Note that this theoretical value assumes that the foam is homogeneous and invariant by translation in both directions, so that at yield stain T1s occur simultaneously and uniformly in the hexagonal foam \cite{Princen_1983, khan_rheology_1986}.

\begin{figure}[h]
  \includegraphics[width=0.9\columnwidth]{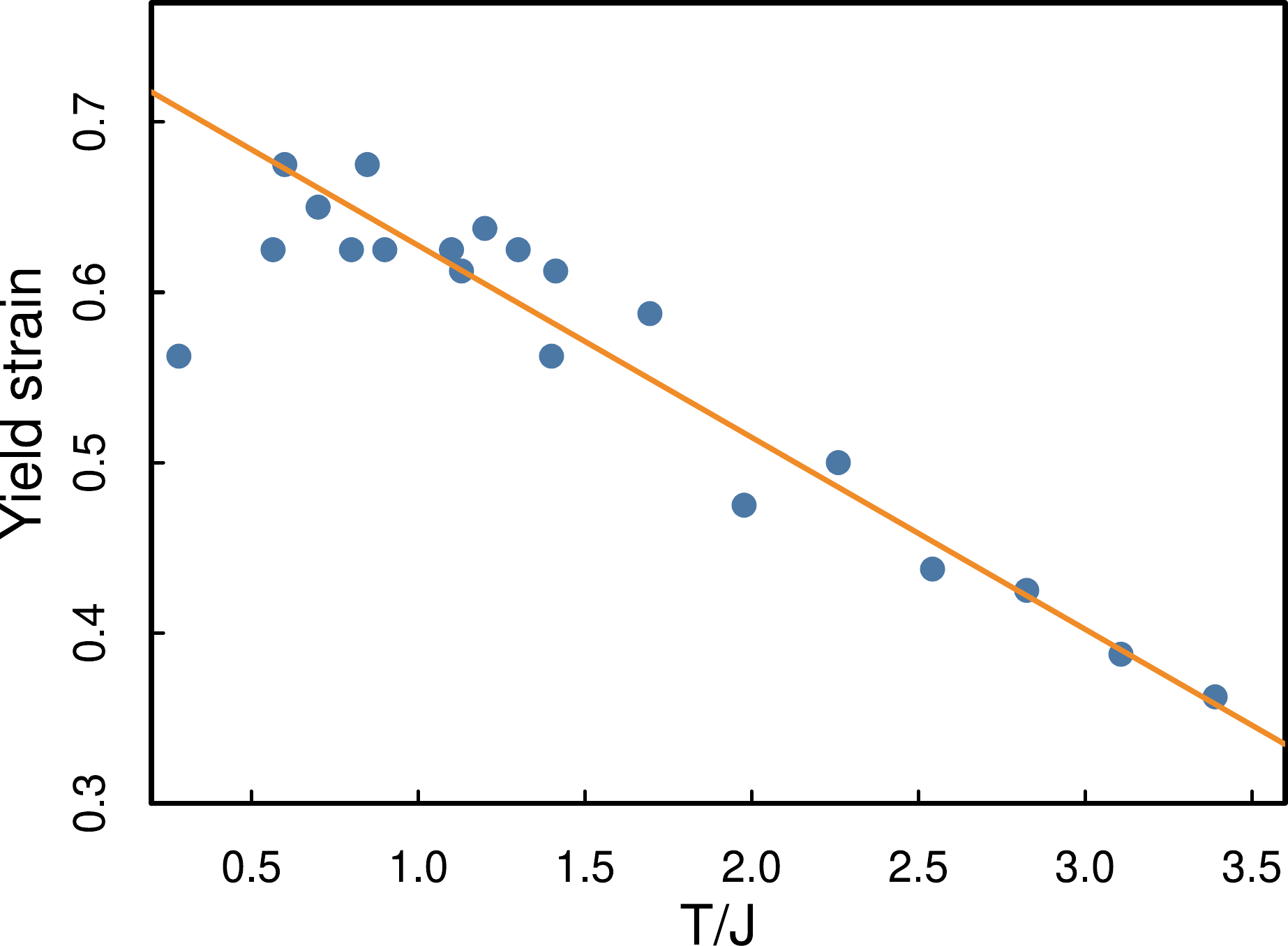}
  \caption{Yield strain as a function of the reduced temperature.}
  \label{fig:yield-strain}
\end{figure}

As for the shear modulus, we study the evolution of the yield strain with temperature and extrapolate to zero temperature to circumvent any effect of the underlying lattice anisotropy  (Fig. \ref{fig:yield-strain}). Extrapolation leads to a yield strain value $0.74$ at zero temperature, which is significantly lower than the theoretical value. The cause of this discrepancy is the presence of the walls: in our simulations, as well as in real foams, the presence of these walls breaks the translational invariance in the $y$ direction, because films meet the walls at right angles \cite{cantat2018foams}.  As a consequence, films in the vicinity of the walls are smaller than in the bulk, as this can be seen in Fig. \ref{fig:shear-bands}, 
and the structure then relaxes through a line of T1s in the vicinity of one of the two walls at strain lower than the theoretical yield value. 



\begin{figure}[h]
	\includegraphics[width=0.9\columnwidth]{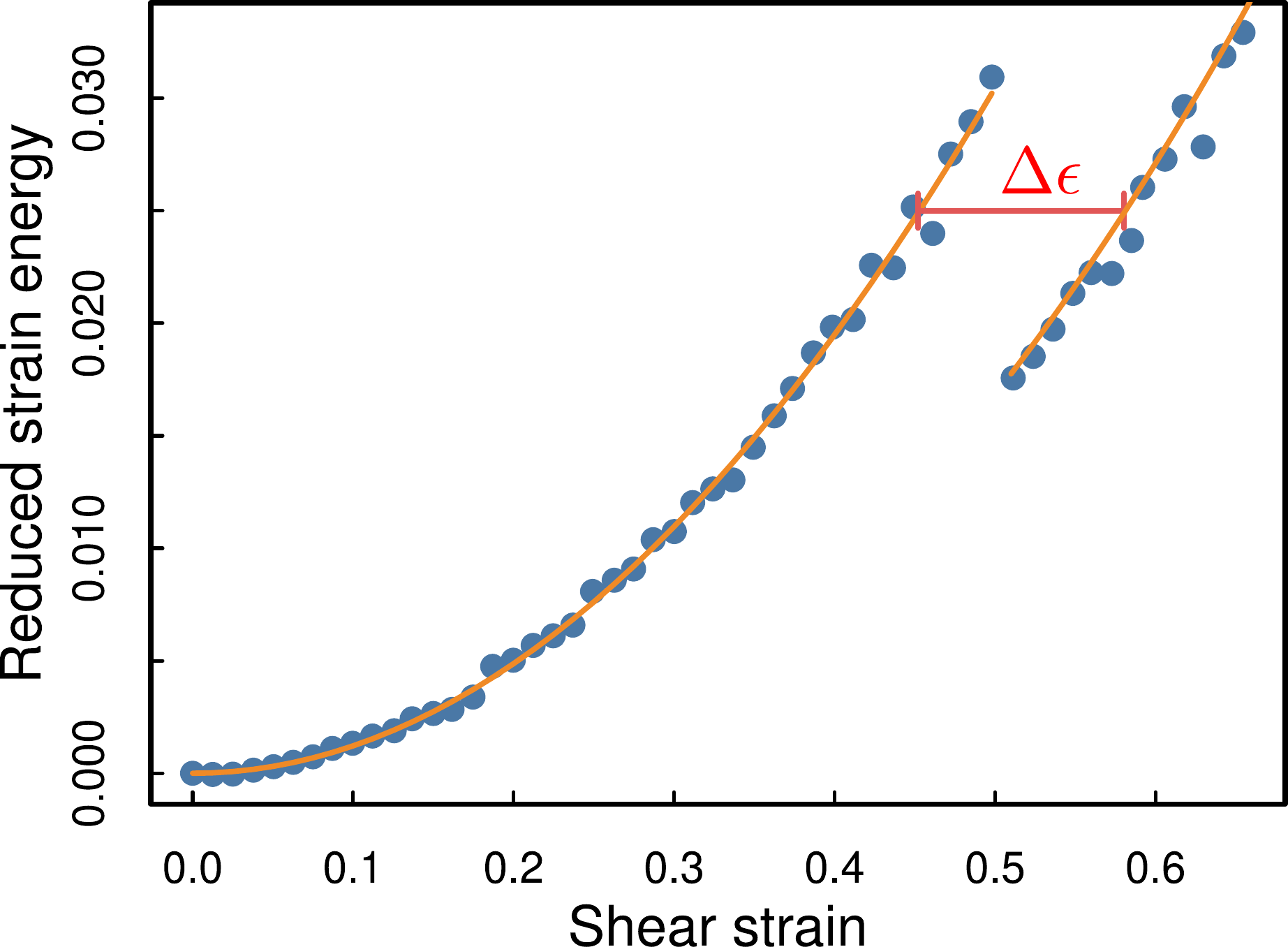}
	\caption{Reduced strain energy $\mathcal E(\epsilon)/ \mathcal E_0-1$ as a function of the shear strain $\epsilon$ (Eq. \ref{eq:energy-strain}). The first orange curve is a quadratic fit of the simulation points below the yield strain of $0.51$. The second curve is identical but offset to the right by a quantity $\Delta \epsilon = 2 / 9\sqrt{3}$.}
	\label{fig:energy-shear-band}
\end{figure}

Because our system is periodic in the $x$ direction, a shear band does not create any topological defect, it just changes the neighbouring of the bubbles that belong to the two rows that slide with respect to each other.
Therefore, the configuration right after the shear banding is still an hexagonal foam, but with a lower effective strain. This strain is actually lowered by a fixed amount: for a system of $N$ rows of bubbles, a translation of $1$ bubble to the side changes the strain by an amount $\Delta \epsilon = 2 / N \sqrt{3}$.
Fig.~\ref{fig:energy-shear-band} shows the reduced strain energy as a function of the shear strain, for a reduced temperature $T/J=2.26$. The abrupt drop in energy corresponds to a shear band.
The first curve on the graph is a quadratic fit of the simulation points below the yield strain of $0.51$. The second curve corresponds to the same parabola, but offset to the right by a quantity $\Delta \epsilon = 2 / 9\sqrt{3}$ (our simulation box contains $10$ rows, two of which are half rows). This second curve has no fitting parameter, and shows that after a shear band the system still has the same shear modulus. 

\subsection{Polydisperse foam}

\begin{figure}[h]
  \includegraphics[width=0.9\columnwidth]{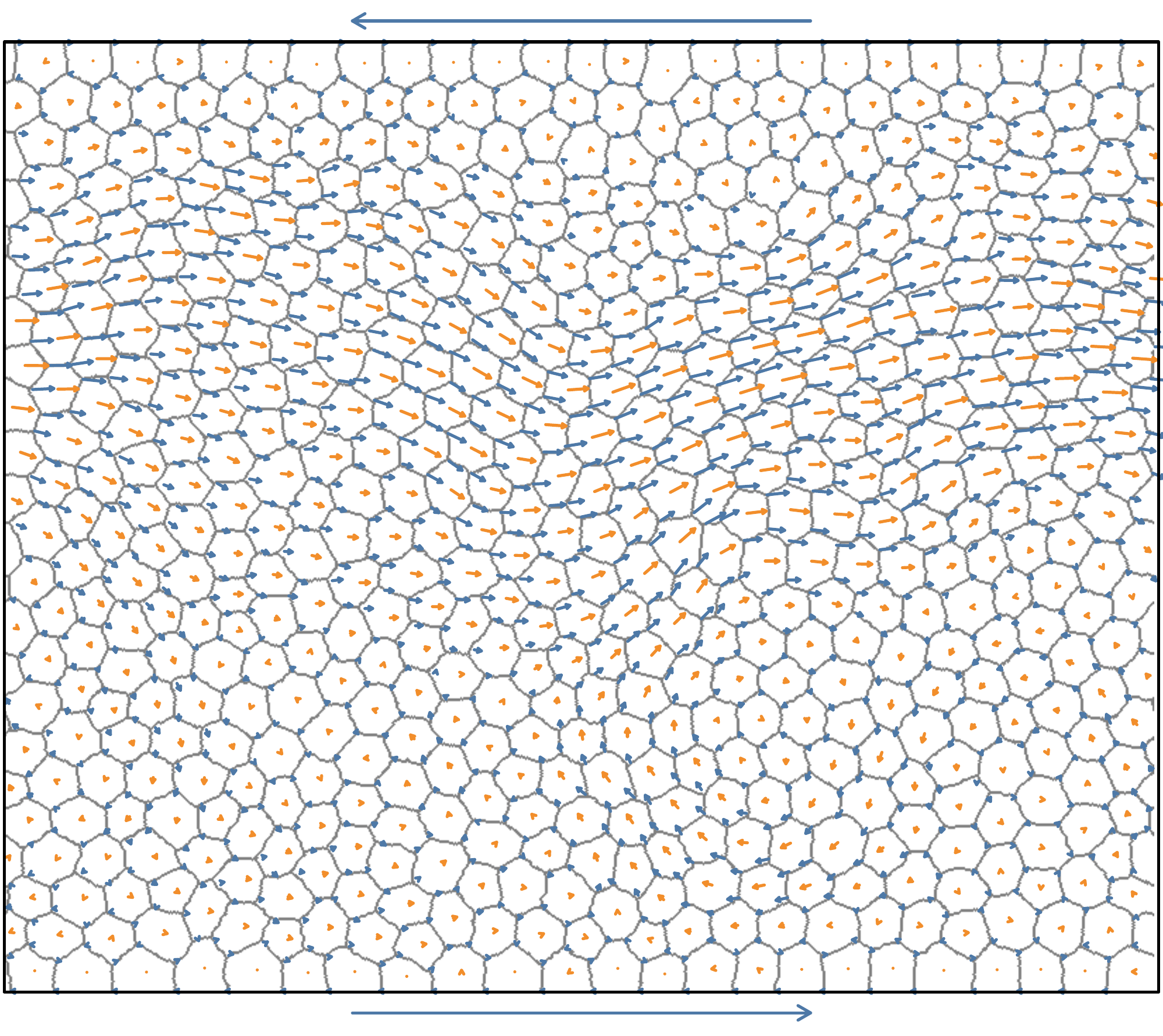}
  \caption{Non-affine component of the displacement field of the bubble centers (in red) and threefold junctions between films (in blue) for the polydisperse foam under quasistatic shear (image corresponds to strain value $\epsilon=0.10$).}
  \label{fig:na-poly}
\end{figure}

We now test our method with a polydisperse, disordered foam. Structural disorder is known to affect the mechanical properties of foams \cite{cox_shear_2006, Hohler_2005}, so it is important to check that our method allows to detect the effect of structural disorder on the mechanical response of a 2D foam.
As for the monodisperse case, we first plot the non-affine component of the displacement field (see Fig. \ref{fig:na-poly}).
In contrast with the regular case, both the bubble centers and the threefold film junctions have a strong non-affine component.

\begin{figure}[h]
  \includegraphics[width=0.9\columnwidth]{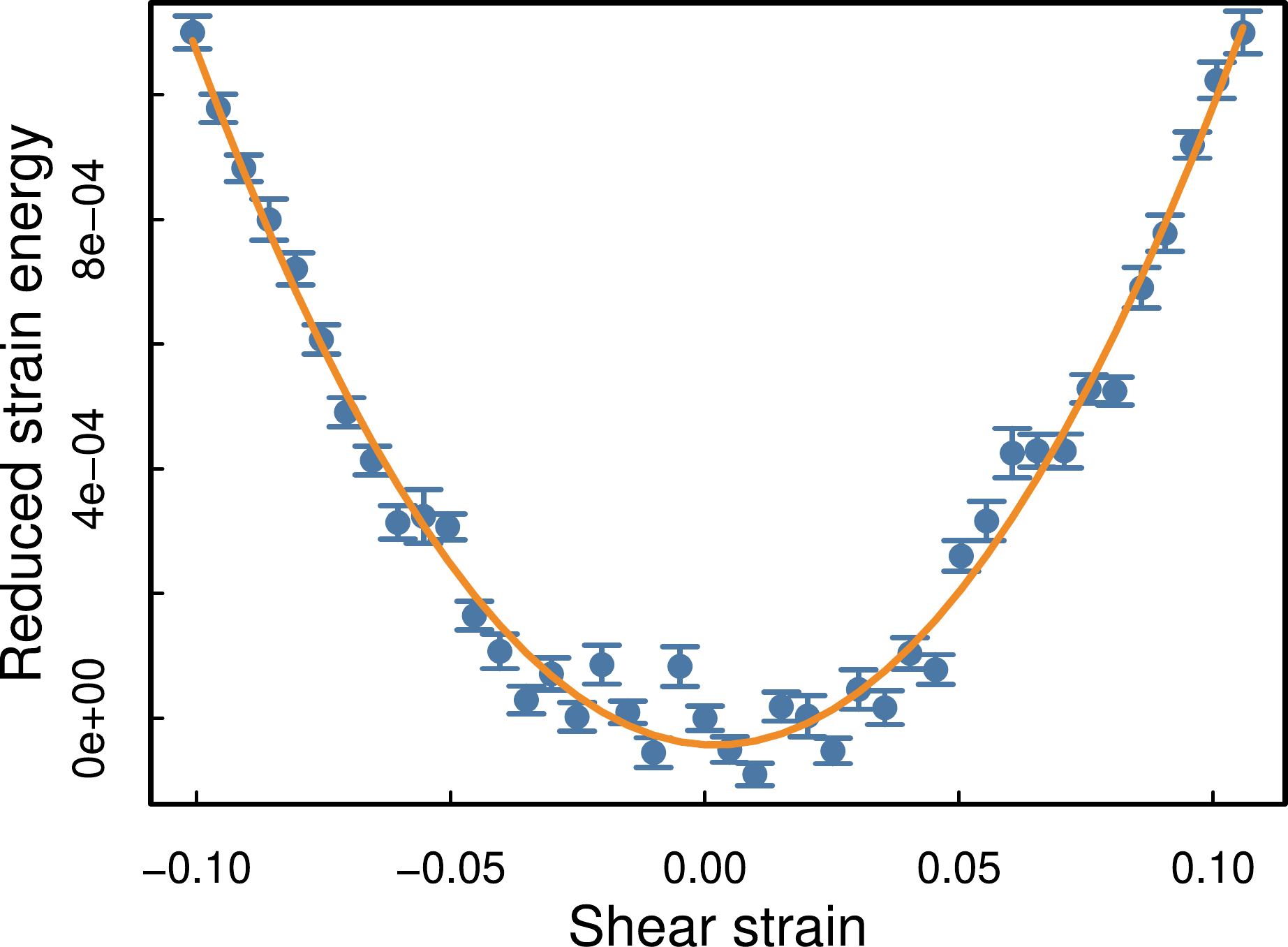}
  \caption{Reduced strain energy $\mathcal E(\epsilon)/ \mathcal E_0-1$ as a function of the shear strain $\epsilon$ (Eq. \ref{eq:energy-strain}). The orange curve is a quadratic fit.}
  \label{fig:energy-shear-poly}
\end{figure}
For the polydisperse foam, shearing was this time done along both  $+x$ and $-x$ directions, as shown on Fig.~\ref{fig:energy-shear-poly}. This setup improves the accuracy of the quadratic fit used to obtain the shear modulus, and allows us to check that there is no residual stress in the initial state 

\begin{figure}[h]
  \includegraphics[width=0.9\columnwidth]{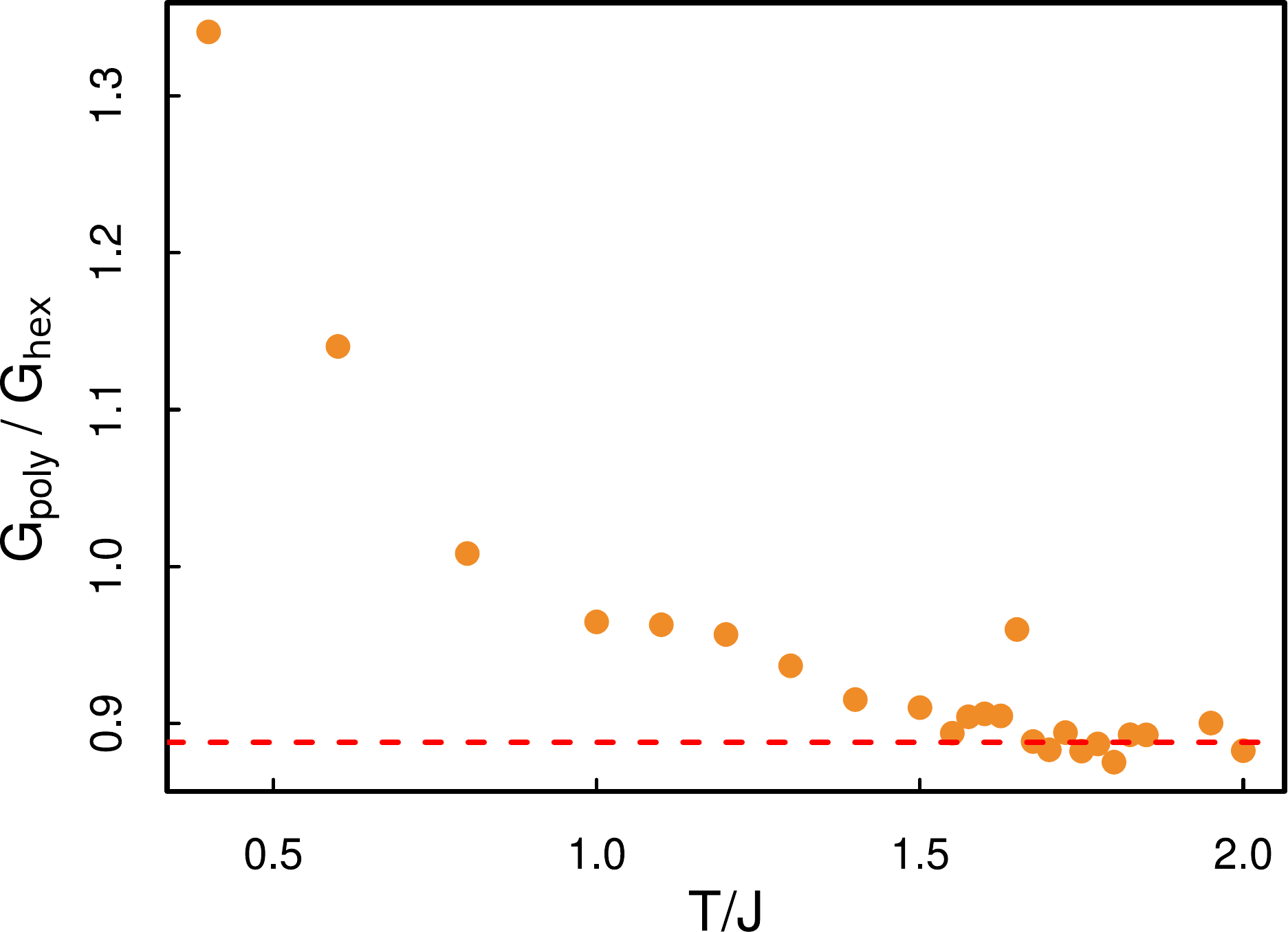}
  \caption{Shear modulus of the polydisperse foam, normalized by the shear modulus of regular hexagonal foam with same mean bubble area and at same temperature, as a function of the reduced temperature $T/J$. The dashed red line represents the average value from simulations with $T/J > 1.5$.}
  \label{fig:G-T-poly}
\end{figure}

Once again, the shear modulus is measured at different temperatures, and then its zero temperature value is extrapolated from linear fit.
For each temperature, the temperature-dependent modulus $G_{\text{poly}}$ was obtained from a quadratic fit of the energy. This modulus was then divided by the shear modulus of the hexagonal foam $G_{\text{hex}}$ at same temperature and with same mean bubble area ($A_0=1000$ pixels$^2$). Note that the ratio $G_{\text{poly}}/G_{\text{hex}}$ is then independent of $z$. 
Fig.~\ref{fig:G-T-poly} shows that this normalized modulus converges to a constant value as temperature is increased. This plateau value is $\sim 89\%$ of the value found for perfect hexagons in our simulation. This value is consistent with those obtained by Cox \& Whittick \cite{cox_shear_2006} with the Surface Evolver program \cite{Brakke_1992}, for foams with similar values of $\Delta A/\langle A\rangle$ and $\Delta n/\langle n \rangle$.


\section{Conclusion}
Numerical simulations are valuable tools to investigate the relationship between the mechanical response and the microscopic details of a cellular material, such as a foam or a biological tissue. 
 The cellular Potts model is a standard numerical modeling tool of multicellular systems, with various applications ranging from foam coarsening to collective behaviors of biological cells. We have shown that quasistatic strain can easily be implemented in CPM, and checked
the accuracy of our method by analyzing the shear strain of 2D foams. For ordered foams, the shear modulus obtained numerically agrees well with the theoretical expression, and bubble centers follow affine displacement as expected. For disordered foams, bubble centers do not follow affine displacement and the shear modulus is found to be lower than for ordered foam with same average bubble area. We obtained good agreement with Surface Evolver simulations, another popular numerical model of multicellular systems. Systematic characterization of the effect of disorder will be investigated in a next study.
More generally, the implementation of quasistatic strain into CPM provides a versatile numerical tool to investigate the interplay between rheological behavior and additional structural changes that take place in cellular systems.

\acknowledgments
ANR (Agence Nationale de la Recherche) and CGI (Commissariat à l’Investissement d’Avenir) are gratefully acknowledged for their financial support of this work through Labex SEAM (Science and Engineering for Advanced Materials and devices), ANR-10-LABX-0096 and ANR-18-IDEX-0001.


\begin{thebibliography}{0}%
\makeatletter
\providecommand \@ifxundefined [1]{%
 \@ifx{#1\undefined}
}%
\providecommand \@ifnum [1]{%
 \ifnum #1\expandafter \@firstoftwo
 \else \expandafter \@secondoftwo
 \fi
}%
\providecommand \@ifx [1]{%
 \ifx #1\expandafter \@firstoftwo
 \else \expandafter \@secondoftwo
 \fi
}%
\providecommand \natexlab [1]{#1}%
\providecommand \enquote  [1]{``#1''}%
\providecommand \bibnamefont  [1]{#1}%
\providecommand \bibfnamefont [1]{#1}%
\providecommand \citenamefont [1]{#1}%
\providecommand \href@noop [0]{\@secondoftwo}%
\providecommand \href [0]{\begingroup \@sanitize@url \@href}%
\providecommand \@href[1]{\@@startlink{#1}\@@href}%
\providecommand \@@href[1]{\endgroup#1\@@endlink}%
\providecommand \@sanitize@url [0]{\catcode `\\12\catcode `\$12\catcode
  `\&12\catcode `\#12\catcode `\^12\catcode `\_12\catcode `\%12\relax}%
\providecommand \@@startlink[1]{}%
\providecommand \@@endlink[0]{}%
\providecommand \url  [0]{\begingroup\@sanitize@url \@url }%
\providecommand \@url [1]{\endgroup\@href {#1}{\urlprefix }}%
\providecommand \urlprefix  [0]{URL }%
\providecommand \Eprint [0]{\href }%
\providecommand \doibase [0]{http://dx.doi.org/}%
\providecommand \selectlanguage [0]{\@gobble}%
\providecommand \bibinfo  [0]{\@secondoftwo}%
\providecommand \bibfield  [0]{\@secondoftwo}%
\providecommand \translation [1]{[#1]}%
\providecommand \BibitemOpen [0]{}%
\providecommand \bibitemStop [0]{}%
\providecommand \bibitemNoStop [0]{.\EOS\space}%
\providecommand \EOS [0]{\spacefactor3000\relax}%
\providecommand \BibitemShut  [1]{\csname bibitem#1\endcsname}%
\let\auto@bib@innerbib\@empty
\end{thebibliography}%


\begin{thebibliography}{27}
\expandafter\ifx\csname natexlab\endcsname\relax\def\natexlab#1{#1}\fi
\expandafter\ifx\csname bibnamefont\endcsname\relax
  \def\bibnamefont#1{#1}\fi
\expandafter\ifx\csname bibfnamefont\endcsname\relax
  \def\bibfnamefont#1{#1}\fi
\expandafter\ifx\csname citenamefont\endcsname\relax
  \def\citenamefont#1{#1}\fi
\expandafter\ifx\csname url\endcsname\relax
  \def\url#1{\texttt{#1}}\fi
\expandafter\ifx\csname urlprefix\endcsname\relax\def\urlprefix{URL }\fi
\providecommand{\bibinfo}[2]{#2}
\providecommand{\eprint}[2][]{\url{#2}}

\bibitem[{\citenamefont{Tlili et~al.}(2015)\citenamefont{Tlili, Gay, Graner,
  Marcq, Molino, and Saramito}}]{tlili_2015}
\bibinfo{author}{\bibfnamefont{S.}~\bibnamefont{Tlili}},
  \bibinfo{author}{\bibfnamefont{C.}~\bibnamefont{Gay}},
  \bibinfo{author}{\bibfnamefont{F.}~\bibnamefont{Graner}},
  \bibinfo{author}{\bibfnamefont{P.}~\bibnamefont{Marcq}},
  \bibinfo{author}{\bibfnamefont{F.}~\bibnamefont{Molino}}, \bibnamefont{and}
  \bibinfo{author}{\bibfnamefont{P.}~\bibnamefont{Saramito}},
  \bibinfo{journal}{The European Physical Journal E}
  \textbf{\bibinfo{volume}{38}}, \bibinfo{pages}{33} (\bibinfo{year}{2015}).

\bibitem[{\citenamefont{Weaire}(2008)}]{Weaire_2008}
\bibinfo{author}{\bibfnamefont{D.}~\bibnamefont{Weaire}},
  \bibinfo{journal}{Current Opinion in Colloid and Interface Science}
  \textbf{\bibinfo{volume}{13}}, \bibinfo{pages}{171} (\bibinfo{year}{2008}),
  ISSN \bibinfo{issn}{1359-0294}.

\bibitem[{\citenamefont{Dollet and Raufaste}(2014)}]{Dollet_2014}
\bibinfo{author}{\bibfnamefont{B.}~\bibnamefont{Dollet}} \bibnamefont{and}
  \bibinfo{author}{\bibfnamefont{C.}~\bibnamefont{Raufaste}},
  \bibinfo{journal}{Comptes Rendus Physique} \textbf{\bibinfo{volume}{15}},
  \bibinfo{pages}{731} (\bibinfo{year}{2014}), ISSN \bibinfo{issn}{1631-0705},
  \bibinfo{note}{liquid and solid foams / Mousses liquides et solides}.

\bibitem[{\citenamefont{G\'eminard et~al.}(2018)\citenamefont{G\'eminard,
  Pastenes, and Melo}}]{Geminard_2018}
\bibinfo{author}{\bibfnamefont{J.-C.} \bibnamefont{G\'eminard}},
  \bibinfo{author}{\bibfnamefont{J.~C.} \bibnamefont{Pastenes}},
  \bibnamefont{and} \bibinfo{author}{\bibfnamefont{F.}~\bibnamefont{Melo}},
  \bibinfo{journal}{Phys. Rev. E} \textbf{\bibinfo{volume}{97}},
  \bibinfo{pages}{042601} (\bibinfo{year}{2018}).

\bibitem[{\citenamefont{Kabla and Debr\'egeas}(2003)}]{Kabla_2003}
\bibinfo{author}{\bibfnamefont{A.}~\bibnamefont{Kabla}} \bibnamefont{and}
  \bibinfo{author}{\bibfnamefont{G.}~\bibnamefont{Debr\'egeas}},
  \bibinfo{journal}{Phys. Rev. Lett.} \textbf{\bibinfo{volume}{90}},
  \bibinfo{pages}{258303} (\bibinfo{year}{2003}).

\bibitem[{\citenamefont{Cox and Whittick}(2006)}]{cox_shear_2006}
\bibinfo{author}{\bibfnamefont{S.~J.} \bibnamefont{Cox}} \bibnamefont{and}
  \bibinfo{author}{\bibfnamefont{E.~L.} \bibnamefont{Whittick}},
  \bibinfo{journal}{The European Physical Journal E}
  \textbf{\bibinfo{volume}{21}}, \bibinfo{pages}{49} (\bibinfo{year}{2006}),
  ISSN \bibinfo{issn}{1292-8941, 1292-895X}.

\bibitem[{\citenamefont{Guyot et~al.}(2019)\citenamefont{Guyot, Kraynik,
  Reinelt, and Cohen-Addad}}]{Guyot_2019}
\bibinfo{author}{\bibfnamefont{P.}~\bibnamefont{Guyot}},
  \bibinfo{author}{\bibfnamefont{A.~M.} \bibnamefont{Kraynik}},
  \bibinfo{author}{\bibfnamefont{D.}~\bibnamefont{Reinelt}}, \bibnamefont{and}
  \bibinfo{author}{\bibfnamefont{S.}~\bibnamefont{Cohen-Addad}},
  \bibinfo{journal}{Soft Matter} \textbf{\bibinfo{volume}{15}},
  \bibinfo{pages}{8227} (\bibinfo{year}{2019}).

\bibitem[{\citenamefont{Cantat et~al.}(2018)\citenamefont{Cantat, Cohen-Addad,
  Elias, Hohler, Pitois, Rouyer, and Saint-Jalmes}}]{cantat2018foams}
\bibinfo{author}{\bibfnamefont{I.}~\bibnamefont{Cantat}},
  \bibinfo{author}{\bibfnamefont{S.}~\bibnamefont{Cohen-Addad}},
  \bibinfo{author}{\bibfnamefont{F.}~\bibnamefont{Elias}},
  \bibinfo{author}{\bibfnamefont{R.}~\bibnamefont{Hohler}},
  \bibinfo{author}{\bibfnamefont{O.}~\bibnamefont{Pitois}},
  \bibinfo{author}{\bibfnamefont{F.}~\bibnamefont{Rouyer}}, \bibnamefont{and}
  \bibinfo{author}{\bibfnamefont{A.}~\bibnamefont{Saint-Jalmes}},
  \emph{\bibinfo{title}{Foams: Structure and Dynamics}}
  (\bibinfo{publisher}{Oxford University Press}, \bibinfo{year}{2018}), ISBN
  \bibinfo{isbn}{9780198824336}.

\bibitem[{\citenamefont{Osborne et~al.}(2017)\citenamefont{Osborne, Fletcher,
  Pitt-Francis, Maini, and Gavaghan}}]{Osborne_2017}
\bibinfo{author}{\bibfnamefont{J.~M.} \bibnamefont{Osborne}},
  \bibinfo{author}{\bibfnamefont{A.~G.} \bibnamefont{Fletcher}},
  \bibinfo{author}{\bibfnamefont{J.~M.} \bibnamefont{Pitt-Francis}},
  \bibinfo{author}{\bibfnamefont{P.~K.} \bibnamefont{Maini}}, \bibnamefont{and}
  \bibinfo{author}{\bibfnamefont{D.~J.} \bibnamefont{Gavaghan}},
  \bibinfo{journal}{PLOS Computational Biology} \textbf{\bibinfo{volume}{13}},
  \bibinfo{pages}{1} (\bibinfo{year}{2017}).

\bibitem[{\citenamefont{Jiang et~al.}(1999)\citenamefont{Jiang, Swart, Saxena,
  Asipauskas, and Glazier}}]{Jiang_1999}
\bibinfo{author}{\bibfnamefont{Y.}~\bibnamefont{Jiang}},
  \bibinfo{author}{\bibfnamefont{P.~J.} \bibnamefont{Swart}},
  \bibinfo{author}{\bibfnamefont{A.}~\bibnamefont{Saxena}},
  \bibinfo{author}{\bibfnamefont{M.}~\bibnamefont{Asipauskas}},
  \bibnamefont{and} \bibinfo{author}{\bibfnamefont{J.~A.}
  \bibnamefont{Glazier}}, \bibinfo{journal}{Physical Review E}
  \textbf{\bibinfo{volume}{59}}, \bibinfo{pages}{5819} (\bibinfo{year}{1999}).

\bibitem[{\citenamefont{Raufaste et~al.}(2007)\citenamefont{Raufaste, Dollet,
  Cox, Jiang, and Graner}}]{Raufaste_2007}
\bibinfo{author}{\bibfnamefont{C.}~\bibnamefont{Raufaste}},
  \bibinfo{author}{\bibfnamefont{B.}~\bibnamefont{Dollet}},
  \bibinfo{author}{\bibfnamefont{S.}~\bibnamefont{Cox}},
  \bibinfo{author}{\bibfnamefont{Y.}~\bibnamefont{Jiang}}, \bibnamefont{and}
  \bibinfo{author}{\bibfnamefont{F.}~\bibnamefont{Graner}},
  \bibinfo{journal}{The European Physical Journal E}
  \textbf{\bibinfo{volume}{23}}, \bibinfo{pages}{217} (\bibinfo{year}{2007}).

\bibitem[{\citenamefont{Princen}(1983)}]{Princen_1983}
\bibinfo{author}{\bibfnamefont{H.~M.} \bibnamefont{Princen}},
  \bibinfo{journal}{Journal of Colloid and interface science}
  \textbf{\bibinfo{volume}{91}}, \bibinfo{pages}{160} (\bibinfo{year}{1983}).

\bibitem[{\citenamefont{Khan and Armstrong}(1986)}]{khan_rheology_1986}
\bibinfo{author}{\bibfnamefont{S.}~\bibnamefont{Khan}} \bibnamefont{and}
  \bibinfo{author}{\bibfnamefont{R.}~\bibnamefont{Armstrong}},
  \bibinfo{journal}{Journal of Non-Newtonian Fluid Mechanics}
  \textbf{\bibinfo{volume}{22}}, \bibinfo{pages}{1} (\bibinfo{year}{1986}),
  ISSN \bibinfo{issn}{03770257}.

\bibitem[{\citenamefont{Glazier et~al.}(1990)\citenamefont{Glazier, Anderson,
  and Grest}}]{Glazier1990}
\bibinfo{author}{\bibfnamefont{J.~A.} \bibnamefont{Glazier}},
  \bibinfo{author}{\bibfnamefont{M.~P.} \bibnamefont{Anderson}},
  \bibnamefont{and} \bibinfo{author}{\bibfnamefont{G.~S.} \bibnamefont{Grest}},
  \bibinfo{journal}{Philos. Mag. B} \textbf{\bibinfo{volume}{62}},
  \bibinfo{pages}{615} (\bibinfo{year}{1990}).

\bibitem[{\citenamefont{Hirashima et~al.}(2017)\citenamefont{Hirashima, Rens,
  and Merks}}]{Hirashima_2017}
\bibinfo{author}{\bibfnamefont{T.}~\bibnamefont{Hirashima}},
  \bibinfo{author}{\bibfnamefont{E.~G.} \bibnamefont{Rens}}, \bibnamefont{and}
  \bibinfo{author}{\bibfnamefont{R.~M.~H.} \bibnamefont{Merks}},
  \bibinfo{journal}{Development, Growth \& Differentiation}
  \textbf{\bibinfo{volume}{59}}, \bibinfo{pages}{329} (\bibinfo{year}{2017}).

\bibitem[{\citenamefont{Graner and Glazier}(1992)}]{Graner_1992}
\bibinfo{author}{\bibfnamefont{F.}~\bibnamefont{Graner}} \bibnamefont{and}
  \bibinfo{author}{\bibfnamefont{J.~A.} \bibnamefont{Glazier}},
  \bibinfo{journal}{Physical Review Letters} \textbf{\bibinfo{volume}{69}},
  \bibinfo{pages}{2013} (\bibinfo{year}{1992}).

\bibitem[{\citenamefont{Szab{\'{o}} et~al.}(2010)\citenamefont{Szab{\'{o}},
  \"{U}nnep, M{\'{e}}hes, Twal, Argraves, Cao, and Czir{\'{o}}k}}]{Szabo_2010}
\bibinfo{author}{\bibfnamefont{A.}~\bibnamefont{Szab{\'{o}}}},
  \bibinfo{author}{\bibfnamefont{R.}~\bibnamefont{\"{U}nnep}},
  \bibinfo{author}{\bibfnamefont{E.}~\bibnamefont{M{\'{e}}hes}},
  \bibinfo{author}{\bibfnamefont{W.~O.} \bibnamefont{Twal}},
  \bibinfo{author}{\bibfnamefont{W.~S.} \bibnamefont{Argraves}},
  \bibinfo{author}{\bibfnamefont{Y.}~\bibnamefont{Cao}}, \bibnamefont{and}
  \bibinfo{author}{\bibfnamefont{A.}~\bibnamefont{Czir{\'{o}}k}},
  \bibinfo{journal}{Physical Biology} \textbf{\bibinfo{volume}{7}},
  \bibinfo{pages}{046007} (\bibinfo{year}{2010}).

\bibitem[{\citenamefont{Kabla}(2012)}]{Kabla_2012}
\bibinfo{author}{\bibfnamefont{A.~J.} \bibnamefont{Kabla}},
  \bibinfo{journal}{Journal of The Royal Society Interface} p.
  \bibinfo{pages}{rsif20120448} (\bibinfo{year}{2012}).

\bibitem[{\citenamefont{Durand and Guesnet}(2016)}]{Durand_2016}
\bibinfo{author}{\bibfnamefont{M.}~\bibnamefont{Durand}} \bibnamefont{and}
  \bibinfo{author}{\bibfnamefont{E.}~\bibnamefont{Guesnet}},
  \bibinfo{journal}{Computer Physics Communications}
  \textbf{\bibinfo{volume}{208}}, \bibinfo{pages}{54} (\bibinfo{year}{2016}),
  ISSN \bibinfo{issn}{00104655}.

\bibitem[{\citenamefont{Durand and Stone}(2006)}]{Durand_2006}
\bibinfo{author}{\bibfnamefont{M.}~\bibnamefont{Durand}} \bibnamefont{and}
  \bibinfo{author}{\bibfnamefont{H.~A.} \bibnamefont{Stone}},
  \bibinfo{journal}{Phys. Rev. Lett.} \textbf{\bibinfo{volume}{97}},
  \bibinfo{pages}{226101} (\bibinfo{year}{2006}).

\bibitem[{\citenamefont{Glazier et~al.}(2007)\citenamefont{Glazier, Balter, and
  Poplawski}}]{Glazier2007b}
\bibinfo{author}{\bibfnamefont{J.~A.} \bibnamefont{Glazier}},
  \bibinfo{author}{\bibfnamefont{A.}~\bibnamefont{Balter}}, \bibnamefont{and}
  \bibinfo{author}{\bibfnamefont{N.~J.} \bibnamefont{Poplawski}}, in
  \emph{\bibinfo{booktitle}{Single-Cell-Based Models in Biology and Medicine}},
  edited by \bibinfo{editor}{\bibfnamefont{D.~A. R.~A.}
  \bibnamefont{Anderson}}, \bibinfo{editor}{\bibfnamefont{P.~M. A.~J.}
  \bibnamefont{Chaplain}}, \bibnamefont{and}
  \bibinfo{editor}{\bibfnamefont{D.~K.~A.} \bibnamefont{Rejniak}}
  (\bibinfo{publisher}{Birkhauser Basel}, \bibinfo{year}{2007}), Mathematics
  and Biosciences in Interaction, pp. \bibinfo{pages}{79--106}.

\bibitem[{\citenamefont{Brakke}(1992)}]{Brakke_1992}
\bibinfo{author}{\bibfnamefont{K.~A.} \bibnamefont{Brakke}},
  \bibinfo{journal}{Experimental Mathematics} \textbf{\bibinfo{volume}{1}},
  \bibinfo{pages}{141} (\bibinfo{year}{1992}),
  \eprint{https://doi.org/10.1080/10586458.1992.10504253}.

\bibitem[{\citenamefont{Magno et~al.}(2015)\citenamefont{Magno, Grieneisen, and
  Marée}}]{Magno_2015}
\bibinfo{author}{\bibfnamefont{R.}~\bibnamefont{Magno}},
  \bibinfo{author}{\bibfnamefont{V.~A.} \bibnamefont{Grieneisen}},
  \bibnamefont{and} \bibinfo{author}{\bibfnamefont{A.~F.}
  \bibnamefont{Marée}}, \bibinfo{journal}{BMC Biophysics}
  \textbf{\bibinfo{volume}{8}} (\bibinfo{year}{2015}), ISSN
  \bibinfo{issn}{2046-1682}.

\bibitem[{\citenamefont{Villemot et~al.}(2020)\citenamefont{Villemot,
  Calmettes, and Durand}}]{Villemot_2020}
\bibinfo{author}{\bibfnamefont{F.}~\bibnamefont{Villemot}},
  \bibinfo{author}{\bibfnamefont{A.}~\bibnamefont{Calmettes}},
  \bibnamefont{and} \bibinfo{author}{\bibfnamefont{M.}~\bibnamefont{Durand}},
  \bibinfo{journal}{Soft Matter} \textbf{\bibinfo{volume}{16}},
  \bibinfo{pages}{10358} (\bibinfo{year}{2020}).

\bibitem[{\citenamefont{Käfer et~al.}(2007)\citenamefont{Käfer, Hayashi,
  Marée, Carthew, and Graner}}]{Kafer_2007}
\bibinfo{author}{\bibfnamefont{J.}~\bibnamefont{Käfer}},
  \bibinfo{author}{\bibfnamefont{T.}~\bibnamefont{Hayashi}},
  \bibinfo{author}{\bibfnamefont{A.~F.} \bibnamefont{Marée}},
  \bibinfo{author}{\bibfnamefont{R.~W.} \bibnamefont{Carthew}},
  \bibnamefont{and} \bibinfo{author}{\bibfnamefont{F.}~\bibnamefont{Graner}},
  \bibinfo{journal}{Proceedings of the National Academy of Sciences}
  \textbf{\bibinfo{volume}{104}}, \bibinfo{pages}{18549}
  (\bibinfo{year}{2007}).

\bibitem[{\citenamefont{Marmottant et~al.}(2009)\citenamefont{Marmottant,
  Mgharbel, Käfer, Audren, Rieu, Vial, Van Der~Sanden, Marée, Graner, and
  Delanoë-Ayari}}]{Marmottant_2009}
\bibinfo{author}{\bibfnamefont{P.}~\bibnamefont{Marmottant}},
  \bibinfo{author}{\bibfnamefont{A.}~\bibnamefont{Mgharbel}},
  \bibinfo{author}{\bibfnamefont{J.}~\bibnamefont{Käfer}},
  \bibinfo{author}{\bibfnamefont{B.}~\bibnamefont{Audren}},
  \bibinfo{author}{\bibfnamefont{J.-P.} \bibnamefont{Rieu}},
  \bibinfo{author}{\bibfnamefont{J.-C.} \bibnamefont{Vial}},
  \bibinfo{author}{\bibfnamefont{B.}~\bibnamefont{Van Der~Sanden}},
  \bibinfo{author}{\bibfnamefont{A.~F.} \bibnamefont{Marée}},
  \bibinfo{author}{\bibfnamefont{F.}~\bibnamefont{Graner}}, \bibnamefont{and}
  \bibinfo{author}{\bibfnamefont{H.}~\bibnamefont{Delanoë-Ayari}},
  \bibinfo{journal}{Proceedings of the National Academy of Sciences}
  \textbf{\bibinfo{volume}{106}}, \bibinfo{pages}{17271}
  (\bibinfo{year}{2009}).

\bibitem[{\citenamefont{Höhler and Cohen-Addad}(2005)}]{Hohler_2005}
\bibinfo{author}{\bibfnamefont{R.}~\bibnamefont{Höhler}} \bibnamefont{and}
  \bibinfo{author}{\bibfnamefont{S.}~\bibnamefont{Cohen-Addad}},
  \bibinfo{journal}{Journal of Physics: Condensed Matter}
  \textbf{\bibinfo{volume}{17}}, \bibinfo{pages}{R1041} (\bibinfo{year}{2005}).

\end{thebibliography}

\end{document}